\documentclass[a4paper,11pt]{article}
\usepackage{jheppub} % for details on the use of the package, please see the JINST-author-manual
\usepackage{lineno}
\usepackage{amssymb}
\usepackage{amsmath}
\usepackage{color}
\usepackage{xspace}
\usepackage{appendix}
\usepackage{float}
\usepackage{hyperref}
\usepackage{diagbox}
\usepackage{multirow}
\usepackage{multicol}
\usepackage{subfig}
\usepackage{comment}
\usepackage{soul}
\usepackage[compat=1.1.0]{tikz-feynhand}
\allowdisplaybreaks
\arxivnumber{2411.07455}

\title{\boldmath Subtraction of the $t\bar{t}$ contribution in $tW\bar{b}$ production at the one-loop level}

\author[a]{Liang Dong,}
\author[a]{Hai Tao Li,}
\author[a]{Zheng-Yu Li,}
\author[a,b]{Jian Wang}

\affiliation[a]{School of Physics, Shandong University, Jinan, Shandong 250100, China}
\affiliation[b]{Center for High Energy Physics, Peking University, Beijing 100871, China}

\emailAdd{liang.dong@mail.sdu.edu.cn}
\emailAdd{haitao.li@sdu.edu.cn}
\emailAdd{zhengyu.li@mail.sdu.edu.cn}
\emailAdd{j.wang@sdu.edu.cn}

\abstract{
The $tW\bar{b}$ production contributes to the real corrections to the $tW$ cross section.
It would interfere with the top quark pair production, causing difficulties in a clear definition of the $tW{\bar b}$ events. 
The subtraction of the $t\bar{t}$ contributions has been performed in the diagram removal or diagram subtraction schemes for the tree-level processes.
However, these schemes rely on the ability to identify the double resonant diagrams and thus can not be extended to loop diagrams.
We propose a new scheme to subtract the $t\bar{t}$ contributions by power expansion of the squared amplitude in the resonant region.
In order to cancel the infra-red divergences of the loop amplitudes, a widely used method is to introduce the dipole counter-terms, an ingredient in calculations of the full next-to-leading order QCD corrections.
In our scheme, these counter-terms are also power-expanded.
As a proof of principle, we calculate the one-loop correction to the $d\bar{d}\to \bar{b}Wt$ process,
and present the invariant mass distribution of the $W\bar{b}$ system.
}

\begin{document}
\maketitle
\flushbottom

\section{Introduction}
\label{sec:intro} 

The top quark is the heaviest elementary particle in the standard model (SM) and plays an important role in studying the electroweak symmetry breaking and searching for new physics signals.
At the Large Hadron Collider (LHC), the dominant production channel is the top and anti-top quark pair production via strong interaction, which has been used to measure the top quark mass of $m_t=172.5 \pm 0.33 $ GeV~\cite{ATLAS:2024dxp} and to perform global fit of the parton distribution function (PDF) and the strong coupling $\alpha_s$~\cite{Alekhin:2024bhs}.
The single top quark production is another source of top quarks and provides a way to determine the Cabibbo-Kobayashi-Maskawa (CKM) matrix element $V_{tb}$~\cite{CDF:2015gsg}.
Moreover, the two processes are the main backgrounds for many new physics signals.
Therefore, it is necessary to calculate the cross sections of these processes with high precision.

The cross section of top quark pair production has been computed up to next-to-next-to-leading order (NNLO) in QCD~\cite{Czakon:2013goa, Catani:2019iny}.
The $t$-channel and $s$-channel single top quark productions have also been studied at NNLO in QCD~\cite{Berger:2016oht, Campbell:2020fhf, Liu:2018gxa}.
In contrast, the $tW$ associated production is only known at next-to-leading order (NLO) in QCD~\cite{Giele:1995kr, Cao:2008af, Kant:2014oha,Campbell:2005bb}.  
The soft gluon effect has been considered at fixed orders~\cite{Kidonakis:2006bu, Kidonakis:2010ux, Kidonakis:2016sjf, Kidonakis:2021vob} and all orders~\cite{Li:2019dhg}. 
Beyond the fixed-order calculations, the simulation of parton shower for $tW$ process has been implemented in~\cite{Frixione:2008yi, Re:2010bp, Jezo:2016ujg}.

Achieving the complete NNLO QCD correction for $tW$ production is crucial for providing accurate predictions of this process.
The first step has been carried out in the calculation of two-loop amplitudes~\cite{Chen:2021gjv,Long:2021vse,Wang:2022enl,Chen:2022ntw, Chen:2022yni, Chen:2022pdw} only recently. 
The slow progress in the theoretical prediction of the $tW$ mode compared to the other two modes, i.e., the $t$- and $s$-channel single top production, is partially due to the interference with the top quark pair production.
In the real correction to $tW$ production, the process of $q\bar{q}/gg\to tW\bar{b}$ interferes with the process of $q\bar{q}/gg\to t\bar{t}(\to W\bar{b}) $.
The significant resonant contribution makes the perturbative expansion in couplings unreliable and a proper definition of the $tW$ process is required, in view of higher-order corrections.
For the NLO correction to $tW$ production, the interference appears only in tree-level diagrams,
and several methods have been proposed to subtract the $t\bar{t}$ contribution~\cite{Belyaev:1998dn, Tait:1999cf, Campbell:2005bb, Frixione:2008yi, Demartin:2016axk, Frixione:2019fxg}.
In some methods, the kinematic cuts, such as the cuts for the invariant mass of $W\bar{b}$ and for the transverse momentum of the $b$-jet, are imposed to suppress the $t\bar{t}$ contribution.
These cuts are not easy to implement in experiments due to the requirements that the $W$ bosons must be clearly reconstructed and that the identification of $b$-jets is of high efficiency. 
Additionally, parts of the signal events would also be rejected under these cuts.
In the other methods, the double resonant diagrams must be distinguished from the others.
However, it is not possible to make such a distinction at higher orders, as we will demonstrate explicitly below.
In this paper, we propose a scheme to subtract the $t\bar{t}$ contribution from the $tW\bar{b}$ production at the one-loop level,
which can be applied in calculating the virtual-real corrections to $tW$ production.

The rest of this paper is organized as follows.
In section \ref{sec:SubScheme}, we first review the conventional subtraction schemes that have been used in the previous calculations, and then propose a subtraction scheme that is applicable at loop levels.
We present the numerical results for the $d\bar{d}\to \bar{b}Wt$ process in section \ref{sec:result},
and conclude in section \ref{sec:conclusion}.
Some calculation details are collected in the appendices.

\section{Subtraction schemes}
\label{sec:SubScheme}

\subsection{Tree level}
\label{sec:BornSub}

\begin{figure}
    \centering
    \begin{minipage}{0.4\textwidth}
        \centering
        \begin{tikzpicture}
        \begin{feynhand}
            \vertex(ddg)at(0,0);\vertex(ttg)at(1,0);\vertex(d1)at(-1,1){$d$};\vertex(d2)at(-1,-1){$\bar{d}$};\vertex(twb)at(1.5,0.5);\vertex(b)at(2.2,1.2){$\bar{b}$};\vertex(w)at(2.2,-0.2){$W$};\vertex(t)at(2.2,-1.2){$t$};\propag[glu](ddg)to(ttg);\propag[fer](d1)to(ddg);\propag[fer](ddg)to(d2);\propag[fer](b)to(twb);\propag[fer,blue](twb)to(ttg);\propag[fer,blue](ttg)to(t);\propag[bos,red](twb)to(w);
        \end{feynhand}
        \end{tikzpicture}
        \caption*{(a)}
    \end{minipage}
    \begin{minipage}{0.4\textwidth}
        \centering
        \begin{tikzpicture}
        \begin{feynhand}
            \vertex(ddg)at(0,0);\vertex(bbg)at(1,0);\vertex(d1)at(-1,1){$d$};\vertex(d2)at(-1,-1){$\bar{d}$};\vertex(twb)at(1.5,-0.5);\vertex(b)at(2.2,1.2){$\bar{b}$};\vertex(w)at(2.2,0.2){$W$};\vertex(t)at(2.2,-1.2){$t$};\propag[glu](ddg)to(bbg);\propag[fer](d1)to(ddg);\propag[fer](ddg)to(d2);\propag[fer](b)to(bbg);\propag[fer](bbg)to(twb);\propag[fer,blue](twb)to(t);\propag[bos,red](twb)to(w);
        \end{feynhand}
        \end{tikzpicture}
        \caption*{(b)}
    \end{minipage}
    \caption{Leading order Feynman diagrams for $d\bar{d}\to\bar{b}Wt$. Diagrams (a) and (b) correspond to the double ($\mathcal{M}^{(0)}_{2t}$) and single ($\mathcal{M}^{(0)}_{1t}$) resonant contributions, respectively. }
    \label{fig:LO}
\end{figure}

The leading order (LO) squared matrix element for the partonic process $A(p_1)+B(p_2)\to \bar{b}(p_3)+W(p_4)+t(p_5)$ can be expressed as
\begin{align}
    |\mathcal{M}_{tW\bar{b}}|_\text{LO}^2=|\mathcal{M}^{(0)}_{1t}+\mathcal{M}^{(0)}_{2t}|^2=|\mathcal{M}^{(0)}_{1t}|^2+2\,\mathbf{Re} [\mathcal{M}^{(0)}_{1t}\mathcal{M}^{(0)\ast}_{2t}]+|\mathcal{M}^{(0)}_{2t}|^2
    \label{eq:A2_tWb},
\end{align}
where $\mathcal{M}^{(0)}_{1t}$ and $\mathcal{M}_{2t}^{(0)}$ denote the LO amplitudes for the single and double top resonant diagrams, respectively.
Typical Feynman diagrams are shown in figure~\ref{fig:LO}.
The double resonant contribution should be considered as top quark pair production and decay, which has been calculated separately.
Therefore, it needs to be subtracted in calculating the cross section of $tW$ production.

The most simple strategy is to remove the contribution from the diagrams of $\mathcal{M}^{(0)}_{2t}$ in eq. (\ref{eq:A2_tWb}) \cite{Frixione:2008yi}, 
\begin{align}
    \left(|\mathcal{M}_{tW\bar{b}}|^2_\text{LO}\right)_\text{DR1}=|\mathcal{M}^{(0)}_{1t}|^2.
\end{align}
A variant of this strategy has also been proposed \cite{Hollik:2012rc},
\begin{align}
    \left(|\mathcal{M}_{tW\bar{b}}|^2_\text{LO}\right)_\text{DR2}=|\mathcal{M}^{(0)}_{1t}|^2+2\,\mathbf{Re} [\mathcal{M}^{(0)}_{1t}\mathcal{M}^{(0)\ast}_{2t}].
\end{align}
Note that the above interference term may give a negative contribution.
This kind of strategy, called {\it diagram removal} (DR), is easy to implement 
and thus has widely been used in experimental analyses.
However, it breaks gauge invariance \cite{Frixione:2008yi}.

One can also keep all the contributions in eq. (\ref{eq:A2_tWb}) intact,
but construct a subtraction term to cancel the double resonant contribution,
\begin{align}
    \left(|\mathcal{M}_{tW\bar{b}}|^2_\text{LO}\right)_\text{Sub}=|\mathcal{M}^{(0)}_{1t}+\mathcal{M}^{(0)}_{2t}|^2-\mathcal{R}_\text{LO}.
    \label{eq:DSmaster}
\end{align}
The subtraction term $\mathcal{R}_\text{LO}$ must have the same value as the double resonant diagrams 
near the on-shell region, which is defined by the limit $\Delta ( \equiv s_{W\bar{b}}-m_t^2 \equiv (p_3+p_4)^2-m_t^2)\to 0$, and decline fast enough in the off-shell regions.
Schematically, we can write
\begin{align}
    \mathcal{R}_\text{LO}=S(\{p_i\},\{\tilde{p}_i\}) \cdot R_\text{LO}(\{\tilde{p}_i\}),
\end{align}
where the subtraction kernel $R_\text{LO}$ has the same behavior as the amplitude squared of top quark pair production while the $S$ factor equals one in the on-shell region but suppresses the $R_\text{LO}$ contribution away from the on-shell region. 
The $S$ factor depends on the original momenta of the initial- and final-state particles, denoted by $\{p_i\}$, and the reshuffled ones, labeled by $\{\tilde{p}_i\}$.
The reshuffled momenta are used in $R_\text{LO}$ so that the on-shell condition $(\tilde{p}_3+\tilde{p}_4)^2=m_t^2$ and gauge invariance are preserved.
We have performed the momentum reshuffling in the same way as in {\tt MC@NLO}~\cite{Frixione:2008yi} and {\tt POWHEG-BOX}~\cite{Re:2010bp};
see appendix~\ref{sec:MomResh} for details.

A natural choice of the subtraction kernel is \cite{Frixione:2008yi}
\begin{align}
    (R_\text{LO})_\text{1}=\widetilde{|\mathcal{M}^{(0)}_{2t}|^2}\equiv |\mathcal{M}^{(0)}_{2t}|^2\Big|_{\{p_i\}\to \{\tilde{p}_i\}},
    \label{eq:R2t_LO_SK1}
\end{align}
where the tilde means that the quantity is calculated with the reshuffled momenta.
This subtraction scheme is called {\it diagram subtraction} (DS). 
In the limit of $\Delta\to 0$, the two terms in eq. (\ref{eq:DSmaster}) would cancel with each other.
However, they are both divergent in this limit, leading to numerical instability.
It is convenient to add an imaginary part in the denominator
of the resonant propagator to regulate such divergences.
The magnitude of this imaginary part is irrelevant as long as it is much smaller than $\Delta$ in the off-shell regions.
We choose it to be $i m_t\Gamma_t$ as indicated by the Breit-Wigner (BW) distribution \cite{Demartin:2016axk}, where the top quark decay width $\Gamma_t $ has been calculated up to NNNLO in the SM; see refs. \cite{Chen:2023dsi, Chen:2023osm} and references therein.

The suppression factor $S$ can be chosen as the ratio of the BW distributions before and after momentum reshuffling \cite{Frixione:2008yi},
\begin{align}
    S_\text{1} & = \frac{(m_t\Gamma_t)^2}{\Delta^2+(m_t\Gamma_t)^2}.
\end{align}
The above choice satisfies the requirement that $S_1 \to 1 $ as $\Delta \to 0$ and $S_1 \to 0$ as $\Delta \gg m_t\,\Gamma_t$.

Therefore, we can express the LO squared amplitude with the subtraction term in the DS scheme as
\begin{align}
    \left(|\mathcal{M}_{tW\bar{b}}|^2_\text{LO}\right)_\text{DS}
    &=\left[|\mathcal{M}^{(0)}_{1t}+\mathcal{M}^{(0)}_{2t}|^2-S_\text{1}\cdot\left(R_\text{LO}\right)_\text{1}\right]_\text{Reg} \notag\\
    &=|\mathcal{M}^{(0)}_{1t}+\mathcal{M}^{(0)}_{2t}|^2_\text{Reg}-\frac{(m_t\Gamma_t)^2}{\Delta^2+(m_t\Gamma_t)^2}\cdot\frac{\widetilde{N}}{\widetilde{\Delta}^2+(m_t\Gamma_t)^2} \notag\\
    &=|\mathcal{M}^{(0)}_{1t}+\mathcal{M}^{(0)}_{2t}|^2_\text{Reg}-\frac{\widetilde{N}}{\Delta^2+(m_t\Gamma_t)^2},
    \label{eq:M2lo_DS1}
\end{align}
where the subscript ``Reg" denotes that we have added $im_t\Gamma_t$
in the anti-top quark propagator. 
In the second line of eq.~(\ref{eq:M2lo_DS1}), 
we have written explicitly the denominator of the resonant propagator with the coefficient $\widetilde{N}$.
The third line is obtained by applying $\widetilde{\Delta}=0$.

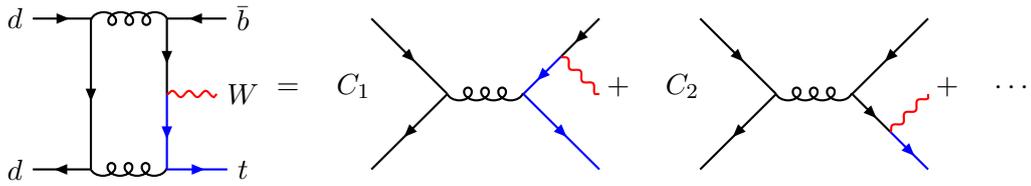
\begin{figure}
    \centering
\begin{align*}
    \begin{tikzpicture}[baseline=0cm]
        \begin{feynhand}
        \vertex(wv)at(0,0);\vertex(bv)at(0,1);\vertex(d1v)at(-1,1);\vertex(d2v)at(-1,-1);\vertex(tv)at(0,-1);\propag[fer](bv)to(wv);\propag[glu](bv)to(d1v);\propag[fer](d1v)to(d2v);\propag[glu](d2v)to(tv);\propag[antfer,blue](tv)to(wv);\vertex(we)at(1,0){$W$};\vertex(be)at(1,1){$\bar{b}$};\vertex(d1e)at(-2,1){$d$};\vertex(d2e)at(-2,-1){$d$};\vertex(te)at(1,-1){$t$};\propag[bos,red](wv)to(we);\propag[fer](d1e)to(d1v);\propag[fer](d2v)to(d2e);\propag[fer](be)to(bv);\propag[fer,blue](tv)to(te);
        \end{feynhand}
    \end{tikzpicture}
    =\quad
    C_1
    \begin{tikzpicture}[baseline=0cm]
        \begin{feynhand}
            \vertex(ddg)at(0,0);\vertex(ttg)at(1,0);\vertex(d1)at(-1,1);\vertex(d2)at(-1,-1);\vertex(twb)at(1.5,0.5);\vertex(b)at(2,1);\vertex(w)at(2,0);\vertex(t)at(2,-1);\propag[glu](ddg)to(ttg);\propag[fer](d1)to(ddg);\propag[fer](ddg)to(d2);\propag[fer](b)to(twb);\propag[fer,blue](twb)to(ttg);\propag[fer,blue](ttg)to(t);\propag[bos,red](twb)to(w);
        \end{feynhand}
    \end{tikzpicture}
    +\quad
    C_2
    \begin{tikzpicture}[baseline=0cm]
        \begin{feynhand}
            \vertex(ddg)at(0,0);\vertex(bbg)at(1,0);\vertex(d1)at(-1,1);\vertex(d2)at(-1,-1);\vertex(twb)at(1.5,-0.5);\vertex(b)at(2,1);\vertex(w)at(2,0);\vertex(t)at(2,-1);\propag[glu](ddg)to(bbg);\propag[fer](d1)to(ddg);\propag[fer](ddg)to(d2);\propag[fer](b)to(bbg);\propag[fer](bbg)to(twb);\propag[fer,blue](twb)to(t);\propag[bos,red](twb)to(w);
        \end{feynhand}
    \end{tikzpicture}
    +\quad\cdots
\end{align*}
    \caption{A typical one-loop resonant diagram for $d\bar{d}\to\bar{b}Wt$. The blue lines represent top quarks and the red wavy lines represent $W$ bosons. After performing the loop integration, the loop diagram reduces to a linear combination of the amplitudes on the right hand of the equation.}
    \label{fig:OnePenta}
\end{figure}

The above methods can not be extended to higher-loop levels
because one can not separate the double resonant diagrams ($\mathcal{M}_{2t}$) from the single resonant ones ($\mathcal{M}_{1t}$).
It is possible that a single loop diagram contains contributions from both the double resonant and the single resonant channels.
For example, the one-loop diagram in figure~\ref{fig:OnePenta} can be considered as a virtual correction to the LO process shown in figure \ref{fig:LO} (a) or figure \ref{fig:LO} (b).
As a result, it is not feasible to remove solely the double resonant diagrams,
and the subtraction term defined in eq. (\ref{eq:R2t_LO_SK1}) can not be well defined at loop levels.

In this work, we propose a new subtraction scheme, dubbed {\it power subtraction} (PS), which can be generalized to loop diagrams.
We note that the squared amplitude can be power expanded around $\Delta=0$ as
\begin{align}
    |\mathcal{M}_{tW\bar{b}}|^2_\text{LO}=\frac{B^{(2)}}{\Delta^2}+\frac{B^{(1)}}{\Delta}+B^{(0)}+\cdots.
\end{align}
In all the coefficients $B^{(i)}$, the relation $s_{W\bar{b}}=m_t^2+\Delta$ has been used.
The first term $\frac{B^{(2)}}{\Delta^2}$ is not equal to $|\mathcal{M}^{(0)}_{2t}|^2$ because the latter contains also  subleading $\Delta$ terms.
We introduce the following subtraction kernel
\begin{align}
    (R_\text{LO})_\text{2}=\frac{\widetilde{B^{(2)}}}{\widetilde{\Delta}^2},
    \label{eq:R2t_LO_SK2}
\end{align}
where $\widetilde{B^{(2)}}$ is obtained from $B^{(2)}$ after momentum reshuffling.
The denominator $\widetilde{\Delta}$ is vanishing in principle.
However, we use it here as a regulator, which will cancel after combination with the suppression factor.
We do {\it not} add an imaginary part in the propagator to avoid numerical instability as above because this would change the infra-red divergence structure of the loop diagrams for on-shell top quark production.
Instead, we define a new suppression factor
as the ratio of the denominators of the resonant propagator before and after momentum reshuffling,
\begin{align}
    S_\text{2}=\frac{\widetilde{\Delta}^2}{\Delta^2}.
    \label{eq:BW3}
\end{align}
The numerator of $S_2$ cancels the pole of $\tilde{\Delta}$ in $ (R_\text{LO})_\text{2}$.
Then the subtracted LO squared amplitude is
\begin{align}
    \left(|\mathcal{M}_{tW\bar{b}}|^2_\text{LO}\right)_\text{PS}
    &=|\mathcal{M}^{(0)}_{1t}+\mathcal{M}^{(0)}_{2t}|^2-S_\text{2}\cdot(R_\text{LO})_\text{2}\notag\\
    &=|\mathcal{M}^{(0)}_{1t}+\mathcal{M}^{(0)}_{2t}|^2-\frac{\widetilde{B^{(2)}}}{\Delta^2}.
    \label{eq:M2lo_DS2}
\end{align}
The two terms will cancel with each other in the limit $\Delta \to 0$.
To ensure numerical stability, we set a cut $|\sqrt{s_{W\bar{b}}}-m_t|/m_t > \delta$ in phase space integration. 
The final cross section should not depend on this cut parameter $\delta$ if it is chosen small enough.
Note that $\widetilde{B^{(2)}}$ coincides with the coefficient $\tilde{N}$ in eq. (\ref{eq:M2lo_DS1}).
Because the power expansion can be performed for all the results of loop diagrams,
this subtraction scheme can be applied beyond the tree level.
The gauge invariance is clearly preserved in this scheme.

\subsection{One-loop level}
\label{sec:loopsub}

When extending the previous scheme to one-loop level, we first expand the squared amplitude around $\Delta = 0$,
\begin{align}
   V_{\text{Ren}}\equiv 2\,\mathbf{Re}[\mathcal{M}^{(0)\ast}\mathcal{M}^{(1)}_\text{Ren}]=\frac{C^{(2)}}{\Delta^2}+\frac{C^{(1)}}{\Delta}+C^{(0)}+\cdots, 
\end{align}
where we have considered the interference between the one-loop and the tree diagrams. 
The first term is what we want to subtract.
We have used the renormalized one-loop amplitude $\mathcal{M}^{(1)}_\text{Ren}$, and thus all the coefficients $C^{(i)}$ contain only infra-red divergences.

In a standard NLO calculation, 
the infra-red divergences of one-loop virtual corrections are canceled with a proper counter-term, such as that in the dipole subtraction formalism~\cite{Catani:1996vz,Catani:2002hc}, i.e.,
the sum 
\begin{align}
    V_{\text{Ren}}+\mathcal{I}_{\text{NLO}}
    \label{eq:nloAct}
\end{align}
is finite.
Here the counter-term can be written as
\begin{align}
    \label{eq:factorize_I}
\mathcal{I}_{\text{NLO}}=\mathbf{I}\otimes|\mathcal{M}_{tW\bar{b}}|^2_{\text{LO}},
\end{align}
where the singular factor $\mathbf{I}$ is factorized from the LO squared amplitude and depends only on the kinematics and color structure of the external particles.
Expanding this counter-term gives
\begin{align}
    \mathcal{I}_{\text{NLO}}=\frac{I^{(2)}}{\Delta^2}+\frac{I^{(1)}}{\Delta}+I^{(0)}+\cdots,
\end{align}
where $I^{(i)}$ are the expansion coefficients.
It is easy to find that each sum of $C^{(i)}+I^{(i)},i=2,1,0$ is finite, i.e., free of $1/\epsilon$ poles.
Now our task is to construct a resonance subtraction term for the combination in eq. (\ref{eq:nloAct}).
To ensure gauge invariance, we need to perform momentum reshuffling, and thus derive the resonance subtraction kernel
\begin{align}
\left(R_\text{V+I}\right)_\text{2}=\frac{\widetilde{C^{(2)}}+\widetilde{I^{(2)}}}{\widetilde{\Delta}^2}
\label{eq:RVI}
\end{align}
with 
\begin{align}
    \widetilde{I^{(2)}}  = I^{(2)}|_{\{p_i\}\to \{\tilde{p}_i\}}.
\end{align}

\begin{figure}
    \centering
    \begin{minipage}{0.3\textwidth}
        \centering
        \begin{tikzpicture}
            \begin{feynhand}
                \vertex(ggg)at(0,0);\vertex(ttg)at(0.9,0);\vertex(b)at(1.8,1);\vertex(w)at(1.8,0);\vertex(t)at(1.8,-1);\vertex(ddg1)at(-0.9,1);\vertex(ddg2)at(-0.9,-1);\vertex(d1)at(-1.8,1);\vertex(d2)at(-1.8,-1);\vertex(twb)at(1.35,0.5);\propag[fer](d1)to(ddg1);\propag[fer](ddg1)to(ddg2);\propag[fer](ddg2)to(d2);\propag[fer](b)to(twb);\propag[fer,blue](twb)to(ttg);\propag[fer,blue](ttg)to(t);\propag[glu](ggg)to(ddg1);\propag[glu](ggg)to(ddg2);\propag[glu](ggg)to(ttg);\propag[bos,red](twb)to(w);
            \end{feynhand}
        \end{tikzpicture}
        \caption*{(a)}
    \end{minipage}
    \begin{minipage}{0.3\textwidth}
        \centering
        \begin{tikzpicture}
            \begin{feynhand}
                \vertex(d2)at(0,0);\vertex(ddg2)at(1.2,0);\vertex(ttg2)at(2.4,0);\vertex(t)at(3.6,0);\vertex(d1)at(0,2);\vertex(ddg1)at(1.2,2);\vertex(ttg1)at(2.4,2);\vertex(b)at(3.6,2);\vertex(twb)at(3,2);\vertex(w)at(3.6,1);\propag[fer](d1)to(ddg1);\propag[fer](ddg1)to(ddg2);\propag[fer](ddg2)to(d2);\propag[fer](b)to(twb);\propag[fer,blue](twb)to(ttg1);\propag[fer,blue](ttg1)to(ttg2);\propag[fer,blue](ttg2)to(t);\propag[glu](ddg1)to(ttg1);\propag[glu](ddg2)to(ttg2);\propag[bos,red](twb)to(w);
            \end{feynhand}
        \end{tikzpicture}
        \caption*{(b)}
    \end{minipage}
    \begin{minipage}{0.3\textwidth}
        \centering
        \begin{tikzpicture}
            \begin{feynhand}
                \vertex(d2)at(0,0);\vertex(ddg2)at(1.2,0);\vertex(ttg2)at(2.4,0);\vertex(t)at(3.6,0);\vertex(d1)at(0,2);\vertex(ddg1)at(1.2,2);\vertex(bbg)at(2.4,2);\vertex(b)at(3.6,2);\vertex(twb)at(2.4,1);\vertex(w)at(3.6,1);\propag[fer](d1)to(ddg1);\propag[fer](ddg1)to(ddg2);\propag[fer](ddg2)to(d2);\propag[fer](b)to(bbg);\propag[fer](bbg)to(twb);\propag[fer,blue](twb)to(ttg2);\propag[fer,blue](ttg2)to(t);\propag[glu](ddg1)to(ttg1);\propag[glu](ddg2)to(ttg2);\propag[bos,red](twb)to(w);
            \end{feynhand}
        \end{tikzpicture}
        \caption*{(c)}
    \end{minipage}
    \caption{Typical one-loop Feynman diagrams contributing to $C^{(2)}$. }
    \label{fig:top_location}
\end{figure}

The coefficient $C^{(2)}$ receives contributions from the loop diagrams with an anti-top quark propagator;
see figure~\ref{fig:top_location} for sample diagrams.
In figure~\ref{fig:top_location}(a), the anti-top quark is not involved in the loop integral and therefore the result can be Taylor expanded around $\Delta = 0$, as at the tree level.
However, the anti-top quark propagator exists inside a loop integral in figure~\ref{fig:top_location}(b) and figure~\ref{fig:top_location}(c).
As a consequence, the result of the squared amplitude contains $\log \Delta$ terms\footnote{The double $\log^2 \Delta$ terms may appear in a single diagram but cancel in the sum of all one-loop diagrams.}, and we can express 
\begin{align}
    C^{(2)} = C^{(2)}_{\textrm{no-log}} +  C^{(2)}_{\rm log} \log \Delta.
\end{align}
Some of the logarithmic terms serve as a regulator for the soft divergence in the diagrams with an on-shell anti-top quark as an external leg, such as the diagram shown in figure~\ref{fig:top_location}(b).
The correspondence between these logarithms and the dimensional divergences is explained in detail in appendix \ref{sec:OffvsOn}.
Then the $\widetilde{C^{(2)}}$ term in eq. (\ref{eq:RVI}) is defined as
\begin{align}
    \widetilde{C^{(2)}}  = C^{(2)}_{\textrm{no-log}}|_{\{p_i\}\to \{\tilde{p}_i\}} + \log\Delta \times C^{(2)}_{\textrm{log}}|_{\{p_i\}\to \{\tilde{p}_i\}} .
\end{align}
Adopting the suppression factor in eq. (\ref{eq:BW3}), we obtain the squared amplitude with both the infra-red and resonant divergences subtracted,
\begin{align}
    \left(|\mathcal{M}_{tW\bar{b}}|^2_\text{V+I}\right)_\text{PS}
    &=V_{\text{Ren}}+\mathcal{I}_\text{NLO}-S_\text{2}\cdot(R_\text{V+I})_\text{2}\notag\\
    &=V_{\text{Ren}}+\mathcal{I}_\text{NLO}-\frac{\widetilde{C^{(2)}}+\widetilde{I^{(2)}}}{\Delta^2}.
    \label{eq:M2loop_DS3}
\end{align}

Our method to construct the resonant contribution is in line with the pole approximation which was developed
in the calculation of the dominant contribution for a process with unstable particles \cite{Stuart:1991xk, Aeppli:1993rs, Denner:2019vbn},
though there are some differences between the two methods.
The resonant contribution is what has to be included in the pole approximation
while it needs to be subtracted in our scheme.
Moreover, we do not include the decay width of the top quark in our formalism.
Therefore the calculation seems simpler and
the momentum reshuffling in the $\log \Delta $ term is different from the on-shell projection used in the pole approximation.

\section{Results for the $d\bar{d}\to \bar{b}Wt$ process}
\label{sec:result}
In this section, we take the $d(p_1)\bar{d}(p_2)\to \bar{b}(p_3)W(p_4)t(p_5)$ process as an example to implement the schemes described in section~\ref{sec:SubScheme}. 
We will present calculation details and numerical results 
at both the tree and one-loop level at the LHC with the center-of-mass energy $\sqrt{S}=14$ TeV.
In numerical evaluation, we use the following input parameters
\begin{gather*}
    \alpha=1/132.16656,\quad \sin^2\theta_W=0.22305189,\\
    m_t=172.5 \text{~GeV},\quad \Gamma_t=1.3\text{~GeV},\quad m_W=80.377 \text{~GeV},
\end{gather*}
where $\alpha$ is the electromagnetic coupling constant and $\theta_W$ is the weak mixing angle. 
Note that the final-state top quark and $W$ boson are stable in our calculation.
The top quark width $\Gamma_t$ is used only as a regulator in the DR2 and DS schemes.
We choose CT18LO~\cite{Yan:2022pzl} and CT18NLO~\cite{Hou:2019qau} PDF sets in the LO and NLO calculations, respectively,
and the associating strong coupling $\alpha_s$ is also used. The renormalization and factorization scales are set to the top quark mass. The CKM matrix is chosen to be diagonal.  We make use of {\tt OpenLoops}~\cite{Buccioni:2019sur} to 
calculate the full amplitudes,
and construct the subtraction terms analytically with the help of {\tt FeynCalc}~\cite{Shtabovenko:2020gxv}.

\subsection{Born cross section}
The LO diagrams of the $d\bar{d}\to \bar{b}Wt$ process are shown in figure~\ref{fig:LO}.
The left diagram in figure~\ref{fig:LO} contains a resonant anti-top quark which decays into $\bar{b}W$. 
Its contribution has to be removed in the DR scheme
or is used to construct a subtraction term in the DS scheme.
In the PS scheme, the subtraction term is given by
\begin{align}
    \frac{\widetilde{B^{(2)}}}{\Delta^2}&=\frac{16\pi^3\alpha\alpha_s^2 C_A C_F}{9m_W^2\tilde{s}_{12}^2\sin^2\theta_W\Delta^2}(m_t^2m_W^2+m_t^4-2m_W^4)\\
    &\left[2m_t^2\tilde{s}_{12}-2m_W^2(\tilde{s}_{12}+2\tilde{s}_{23}+2\tilde{s}_{24})+2m_W^4+\tilde{s}_{12}^2+2(\tilde{s}_{23}+\tilde{s}_{24})(\tilde{s}_{12}+\tilde{s}_{23}+\tilde{s}_{24})\right],\notag
\end{align}
where $C_A=3$ and $C_F=4/3$ in QCD.
The Mandelstam variables are defined by
\begin{align}
    \tilde{s}_{ij}=(\tilde{p}_i + \sigma_{ij} \tilde{p}_j)^2.
\end{align}
The sign $\sigma_{ij}=1$ if the $i$- and $j$-th particles are both initial or both final states; and $\sigma_{ij}=-1$ in the other cases.
After phase space integration and convolution with the PDFs, the results of the cross section in different schemes are presented in table \ref{tab:BornXS}.
Note that, in the PS scheme, we have imposed a cut, $|\sqrt{s_{W\bar{b}}}-m_t|/m_t>\delta$, for numerical stability when implementing the Monte Carlo integration, since large number cancellation appears near the resonant region.  
The value of $\delta$ does not affect the numerical result significantly as long as $\delta \ll 1$. We have checked that in the interval for $\delta$ between $10^{-2}$ and $10^{-6}$ the integration result is stable. For the cross section in the PS scheme in table~\ref{tab:BornXS}, we have used $\delta=0.006$. 

It can be seen that the results in the DR1, DS, and PS schemes are similar in size.
A negative result is obtained in the DR2 scheme because of the contribution from the interference term.
This may be not a problem if one considers this process as a real correction to the $tW$ production, in which much larger positive corrections come from the virtual corrections and the real corrections without $\bar{b}Wt$ final states.
However, given that the $d\bar{d}\to\bar{b}Wt$ process does not require subtraction of infra-red divergences, it is more reasonable to have a positive value for its contribution.

\begin{table}
    \centering
    \begin{tabular}{ccccc}
        \hline
        Schemes & DR1 & DR2  & DS & PS \\ \hline
         $\sigma_\text{LO}$(fb)  & $63.56(1)$  
         & $-45.03(1)$ 
          & $56.7(2)$
          & $56.1(1)$ \\
        \hline
    \end{tabular}
    \caption{Born cross sections for $d\bar{d}\to\bar{b}Wt$ process calculated in different resonance subtraction schemes. The numerical uncertainty affecting the last digit is quoted in parentheses. }
    \label{tab:BornXS}
\end{table}

\begin{figure}
    \centering
    \includegraphics[scale=0.7]{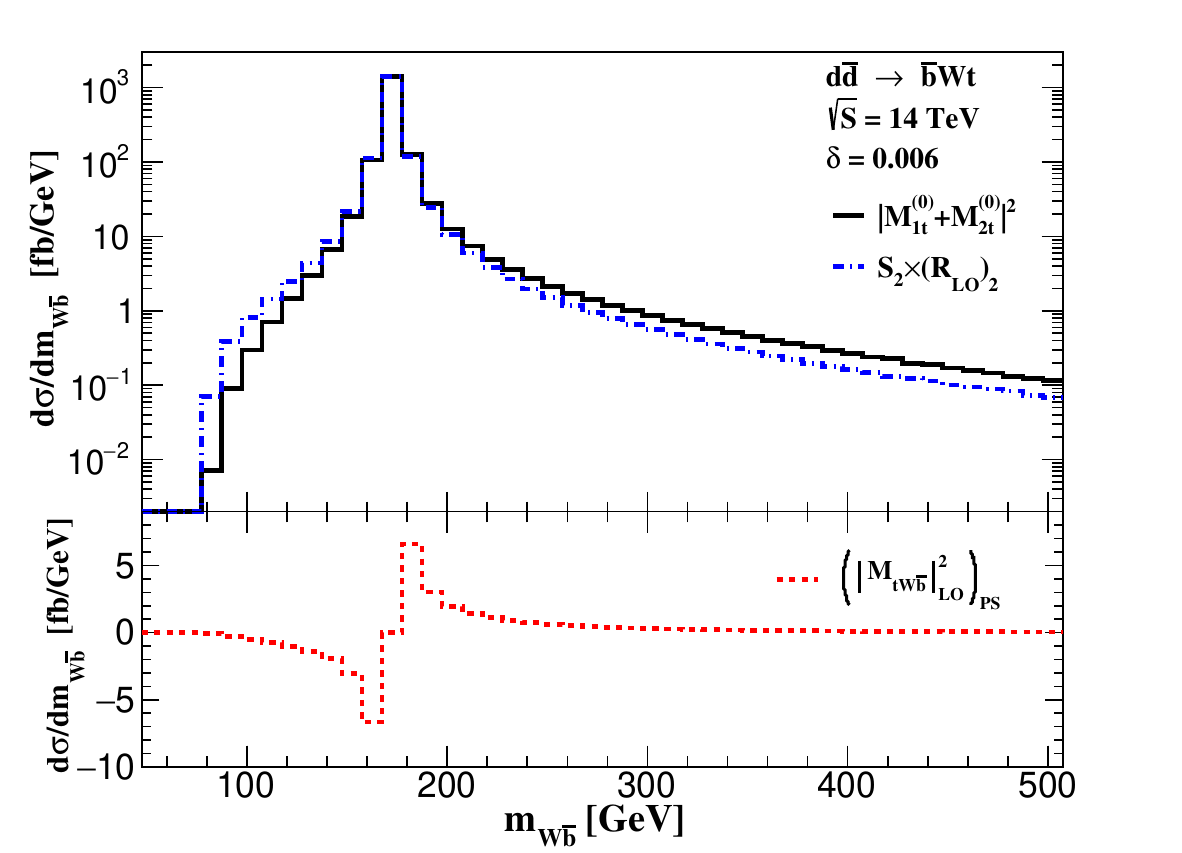}
    \caption{The invariant mass distribution of the $W^-\bar{b}$ system in the LO result of $d\bar{d} \to \bar{b}W^-t$.}
    \label{fig:LO_distrib}
\end{figure}

To see more clearly the subtraction pattern, we show the invariant mass distribution for the $W\bar{b}$ system in figure~\ref{fig:LO_distrib}.
In the bin including the resonance, a cancellation of five digits occurs, illustrating the validity of the subtraction term.
The cross sections in the two adjoining bins drop by an order of magnitude and the cancellation between the cross section and the subtraction term is still obvious.
The differential cross sections in the other bins are even smaller and the subtraction term plays a less significant role.
From the lower panel of figure~\ref{fig:LO_distrib}, we see that the subtracted cross section still contains a single pole of $\Delta$.
The contributions on the two sides of the resonance peak are almost the same but with opposite signs.
The eleven bins around the resonance peak, i.e., $117.5{\rm ~GeV}<m_{W\bar{b}} < 227.5$ GeV contribute less than 1\% of total cross section.
Therefore, these resonant contributions to the differential distributions other than $m_{W\bar{b}}$ would be very small.

\subsection{One-loop corrections}
\label{sec:subterm}

The one-loop diagrams can be separated into three types as shown in figure \ref{fig:top_location}.
The amplitudes of the first two types of diagrams depend on four-point (or less point) scalar integrals,
of which the analytic results can be found in the literature \cite{Ellis:2007qk,Patel:2016fam}.
Therefore it is straightforward to expand the results in a series of $\Delta$.
In order to obtain the analytic result of the pentagon integral, denoted by $I_5$, in the Feynman diagram of the third type (figure \ref{fig:top_location}(c)), we need to firstly reduce it to a linear combination of five box integrals \cite{Bern:1993kr}, 
\begin{align}
    I_5=\sum_{i=1}^5 c_i I_4^{(i)}+\mathcal{O}(\epsilon),
\end{align}
where the coefficients $c_i$ depend on the kinematic variables and masses, and the box integrals $I_4^{(i)}$ can be obtained after removing one of the propagators in the pentagon integral.
Then one can power expand these box integrals analytically.
We find that only the following integral contains $\mathcal{O}(1/\Delta)$ contribution,
\begin{align}
    \int d^Dl \frac{1}{(l^2+i0)[(l-p_3)^2+i0][(l-p_3-p_4)^2-m_t^2+i0][(l-p_1)^2+i0]}
\end{align}
with $D=4-2\epsilon$.

Actually, the expansion of the amplitude in $\Delta$ can be described systematically by using the effective field theory of unstable particles \cite{Beneke:2003xh}.
The idea is that the anti-top quark in a loop diagram can only have soft momentum fluctuation near the resonant region.
As a result, the momentum of the gluon connecting the two partons in the production and decay processes must be soft.
We can extract the leading power contribution just by making use of the eikonal approximation of the gluon interaction.
In other words, we consider such gluons arising from the soft Wilson lines associating with the corresponding partons.
Such soft loop integrals are usually vanishing in QCD amplitudes because they are scaleless.
But they are non-vanishing in our case because the anti-top quark propagator, which introduces a scale of the offshellness, should also be taken into account. 
As an example, when calculating the Feynman diagram in figure \ref{fig:top_location}(c), we only need to evaluate the following soft loop integral 
\begin{align}
    \label{eq:softpentaint}
    \int d^Dl \frac{1}{(l^2+i0)(-2l\cdot p_3+i0)\left(-2l\cdot (p_3+p_4)+\Delta+i0\right)(-2l\cdot p_1+i0)}.
\end{align}
Note that the soft gluon momentum $l^{\mu}$ is of the order $\lambda^2\sim \Delta/m_t^2 $.
This kind of integrals is general for the soft gluon correction in Feynman diagrams with a resonant propagator.
Its analytical result, along with other soft loop integrals, is shown in appendix \ref{sec:integrals}.

Making use of this method, we compute the leading power expansion of the one-loop amplitude and obtain
\begin{align}
    \label{eq:DivInRes}
    C^{(2)}=&\frac{\alpha_s }{2\pi}B^{(2)}\Bigg\{-\frac{3C_F}{\epsilon^2}+\frac{1}{\epsilon}\Bigg[ (C_A-2C_F)\log\left(\frac{\mu^2}{s_{12}}\right)-2(C_A-2C_F)\log\left(\frac{\mu^2}{-s_{13}}\right) \notag\\
    &+(C_A-4C_F)\log\left(\frac{\mu^2}{-s_{23}}\right)+(C_A-4C_F)\log\left(\frac{m_t \mu}{-s_{15}+m_t^2}\right) \\
    &
    -2(C_A-2C_F)\log\left(\frac{m_t \mu }{-s_{25}+m_t^2}\right)+(C_A-2C_F)\log\left(\frac{m_t\mu}{s_{35}-m_t^2}\right)
    -\frac{11}{2} C_F \Bigg]\Bigg\}\notag\\
    & +\cdots, \notag
\end{align}
where we have only written the divergent parts explicitly.
We see that the divergent part is proportional to the LO analogue $B^{(2)}$ and does not contain any $\log\Delta$ terms\footnote{It is worth noting that there may appear the logarithmic term $\log\Delta$ in a single diagram.}.
This feature is expected because the sum $C^{(2)} + I^{(2)}$ should be finite and $I^{(2)}$ does not depend on $\Delta$.
In real calculation, it results from color conservation and the structure of soft-collinear divergences. The relevant contribution can be written as
\begin{align}
  & \frac{\alpha_s}{2\pi\epsilon^2} \left(\frac{-\Delta-i0}{\mu\,m_t}\right)^{-2\epsilon}
    \bigg[ 
2{\bf T}_1\cdot {\bf T}_3 
+2{\bf T}_2\cdot {\bf T}_3 
+{\bf T}_3\cdot {\bf T}_{5} 
-{\bf T}_1\cdot {\bf T}_{\bar t,f} 
-{\bf T}_2\cdot {\bf T}_{\bar t,f} 
-{\bf T}_3\cdot {\bf T}_{\bar t,i} 
    \bigg]
    \nonumber\\
  =& 
    \frac{\alpha_s}{2\pi\epsilon^2} \left(\frac{-\Delta-i0}{\mu\,m_t}\right)^{-2\epsilon}
    \bigg[ 
{\bf T}_1\cdot {\bf T}_3 
+{\bf T}_2\cdot {\bf T}_3 
+{\bf T}_3\cdot {\bf T}_{5} 
+{\bf T}_3\cdot {\bf T}_{\bar t,f} 
    \bigg]  =0\,,
\end{align}
where ${\bf T}_i$'s are color operators~\cite{Catani:1996vz} and we have used color conservation ${\bf T}_{\bar t,f}={\bf T}_3=-{\bf T}_{\bar t,i}$ and 
${\bf T}_1+{\bf T}_2+{\bf T}_5+{\bf T}_{\bar t,f}=0$ in the second line.
The coefficient of the color operators in each term is determined by the soft-collinear divergence associating with the corresponding external states; see appendix \ref{sec:integrals} for the results of soft loop integrals.

The ellipsis in eq. (\ref{eq:DivInRes}) denotes the finite piece. The complete form is too lengthy to be shown here.
The $\log\Delta$ terms in the finite piece are also proportional to the LO  $B^{(2)}$ and can be expressed by
\begin{align}~\label{eq:lnDelta}
    2\log\left(\frac{m_t\mu}{|\Delta|}\right)\cdot  \frac{\alpha_s}{2\pi} B^{(2)} & \Bigg[(C_A-2C_F) \
    \left( \log\left(\frac{m_t^2}{s_{35}-m_t^2}\right)-\frac{1+\beta_t^2} {2\beta_t} \log \left( \frac{1-\beta_t}{1+\beta_t} \right)\right)
    \notag\\
    & -2(C_A-2C_F) \left( 
    \log\left(\frac{m_t^2}{-s_{13}}\right) - \log\left(\frac{m_t^2}{2 p_1 \cdot p_{\bar{t}}}\right) 
     \right)
    \notag\\
    &+(C_A-4C_F)\left( \log\left(\frac{m_t^2}{-s_{23}}\right)-\log\left(\frac{m_t^2}{2 p_2 \cdot p_{\bar{t}}}\right)\right) 
    \notag\\
    & +2C_F\log\left(\frac{m_t^2}{2 p_{\bar{t}}\cdot p_3}\right)
    +2C_F\Bigg],
\end{align}
where we have introduced $p_{\bar{t}} = p_3+p_4$ and $\beta_t=\sqrt{1-4m_t^2/s_{12}}$ is the velocity of the top or anti-top quark in the center-of-mass frame in the limit $\Delta \to 0 $. 
The above expression in the bracket is equal to the difference between the infra-red divergences of the matrix elements for $d\bar{d}\to \bar{b}Wt$ and $d\bar{d}\to \bar{t}t$ with $\bar{t}\to \bar{b}W$ normalized by their respective Born results and $\alpha_s/2\pi\epsilon$. 
To be specific, the infra-red divergence for $d\bar{d}\to \bar{b}Wt$ with an off-shell anti-top propagator can be written schematically as  
\begin{align}~\label{eq:offshell}
     I_{d\bar{d}\to \bar{b}Wt}^{\rm div} &= \frac{d_2}{\epsilon^2}+ \frac{d_1}{\epsilon} +\frac{d_s}{\epsilon}\left(\frac{-\Delta-i0}{\mu\,m_t}\right)^{-2\epsilon}
     =\frac{d_2}{\epsilon^2}+ \frac{d_1+d_s}{\epsilon} -2 d_s \log \left(\frac{-\Delta-i0}{\mu\,m_t}\right)\,,
\end{align}
while the infra-red  structure for  $d\bar{d}\to \bar{t}t$ with on-shell anti-top decaying $\bar{t}\to \bar{b}W$ is 
\begin{align}~\label{eq:onshell}
    I_{d\bar{d}\to\bar{t}(\to\bar{b}W)t}^{\rm div} &= \frac{d_2}{\epsilon^2}+ \frac{d_1}{\epsilon}\,.
\end{align}
In the limit $\Delta\to 0$, the two results should coincide, which explains the same coefficients $d_2$ and $d_1$ in the above two equations. 
With small but non-vanishing $\Delta$, expansion in $\epsilon$ as shown in the second equation of (\ref{eq:offshell})  gives rise to the $\log (|\Delta|)$ terms in the finite piece in (\ref{eq:lnDelta}). 
Because of this relation, the gauge invariance of $ \widetilde{C^{(2)}} $ is also obvious.

\begin{figure}[ht]
    \centering
    \includegraphics[scale=0.6]{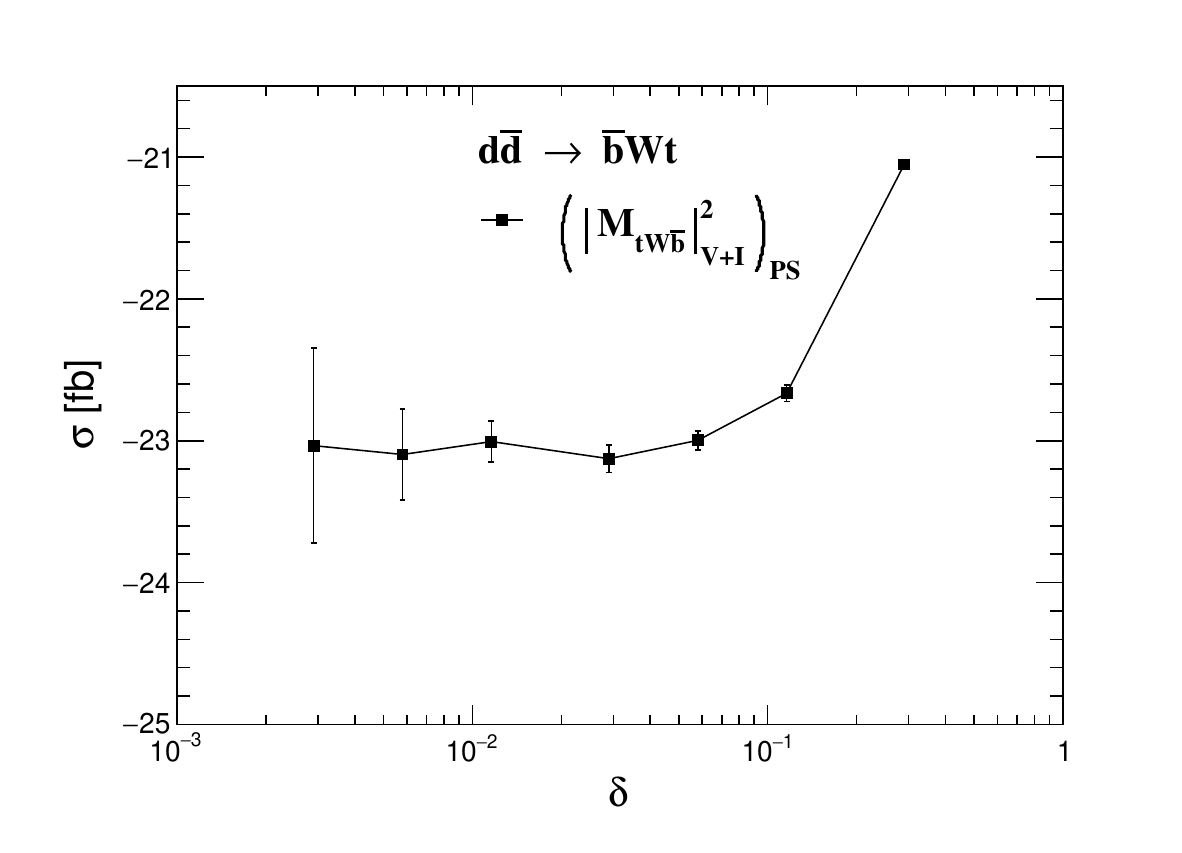}
    \caption{Contributions of the one-loop correction and integrated dipole counter-terms for $d\bar{d}\to\bar{b}Wt$ process in the PS scheme for various values of $\delta$. The error bar represents the numerical uncertainty from Monte Carlo integration.}
    \label{fig:delta}
\end{figure}

We have imposed a phase space cut $|\sqrt{s_{W\bar{b}}}-m_t|/m_t>\delta$ in the Monte Carlo integration.
The one-loop corrections with resonant contribution subtracted for different values of $\delta$ are shown in figure~\ref{fig:delta}.
We find that the results become insensitive to $\delta$ when it becomes smaller than 0.1.
It is ideal that one can take the limit of $\delta \to 0$.
However, this is not possible in practice because
the numerical uncertainties due to Monte Carlo integration grow with decreasing $\delta$.
Given that the central values almost remain the same, 
it is reasonable to choose the result at $\delta = 0.006$.
The LO cross section and the one-loop correction are shown in table \ref{tab:VirtXS}. We see that the virtual correction decreases the LO result by 41\%.

\begin{table}[ht]
    \centering
    \begin{tabular}{cccc}
        \hline
        Scheme & $\sigma_\text{LO}$(fb) & $\sigma_\text{V+I}$(fb) & $\sigma_\text{V+I}/\sigma_\text{LO}$ \\ \hline 
        PS & 56.1(1) & -23.1(3) & -0.41 \\
        \hline
    \end{tabular}
    \caption{LO cross section and contributions of the one-loop correction and integrated dipole counter-terms for the $d\bar{d}\to\bar{b}Wt$ process in the PS scheme.  The numerical uncertainty affecting the last digit is quoted in parentheses. We have chosen $\delta=0.006$.}
    \label{tab:VirtXS}
\end{table}

\begin{figure}
    \centering
    \includegraphics[scale=0.7]{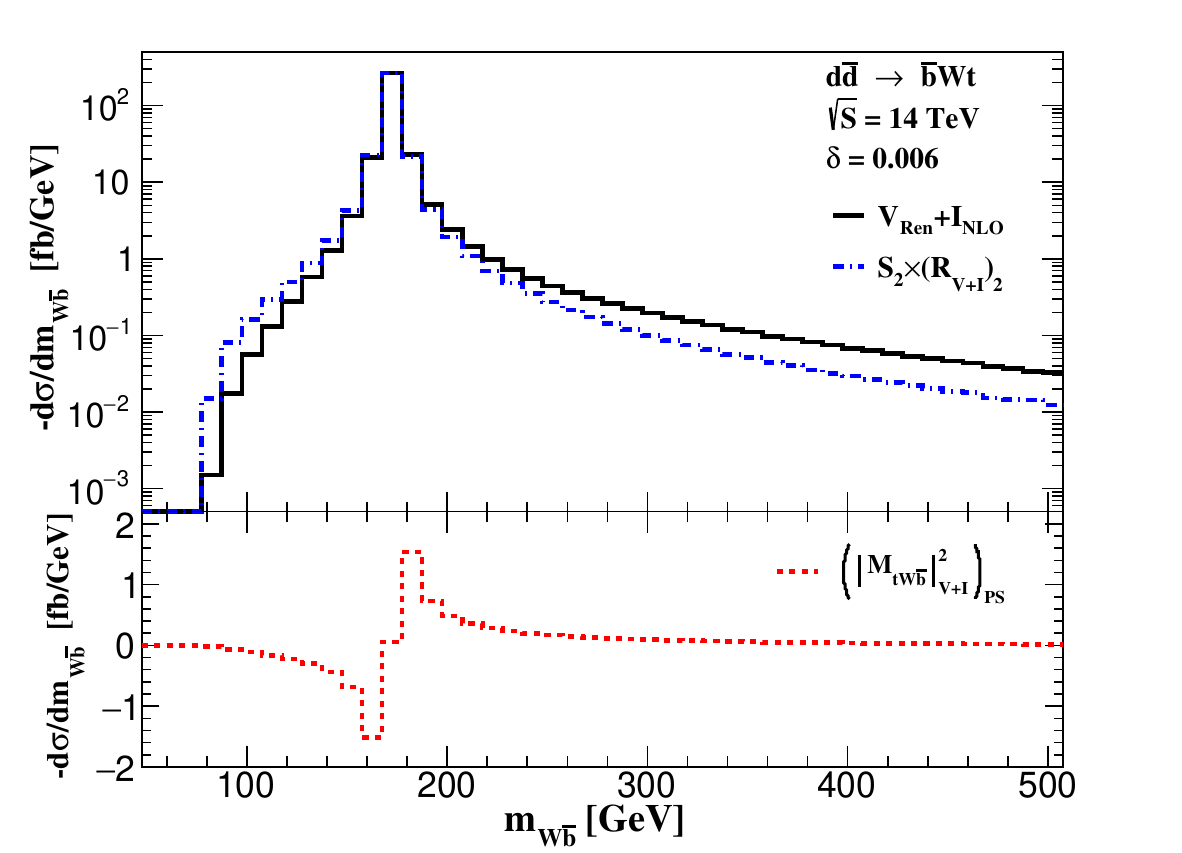}
    \caption{The invariant mass distribution of the $W^-\bar{b}$ system in the one-loop correction to $d\bar{d} \to \bar{b}Wt$. }
    \label{fig:V+I_distrib}
\end{figure}

The invariant mass distribution for the $W\bar{b}$ system is shown in figure~\ref{fig:V+I_distrib}.
A large cancellation can be observed near the onshell region. 
The resonance subtraction term drops fast away from the onshell region, as in the LO.
The two bins adjacent to the resonance peak provide opposite contributions, illustrating the remaining $1/\Delta$ structure in the squared amplitude. 
The contributions from the nine bins around the resonance peak amount to only 10\% of the total cross section, indicating a good cancellation effect of the resonant contributions in the PS scheme at the one-loop level.

\section{Conclusion}
\label{sec:conclusion}

The $tW$ associated production is an important process at the LHC. 
Precise calculation of its cross section requires the subtraction of the $\bar{t}(\to \bar{b}W)t$ events.
In the calculation of NLO corrections, it is sufficient to perform the subtraction for the tree-level process of $pp \to tW\bar{b}$.
Two schemes, i.e., the diagram removal and diagram subtraction, have been widely used in the past.
However, they are based on the ability to recognize the resonant contribution in Feynman diagrams, and thus can not be extended to higher orders,
since there exist loop diagrams which contain both resonant and non-resonant contributions.

We have proposed a new subtraction scheme to address such a problem. 
We first expand the squared amplitudes around the resonant limit of $\Delta\to 0$ in powers of $\Delta$, 
and then subtract the leading power contribution.
Moreover, the expansion is also carried out for the integrated subtraction terms in the formalism of dipole subtraction. 
We applied such a scheme in calculating the one-loop corrections to the $d\bar{d}\to tW\bar{b}$ process.
The effectiveness of this scheme is observed in the invariant mass distribution of the $W\bar{b}$ system.
We believe that this scheme is also applicable to the calculation of the real correction to the $d\bar{d}\to tW\bar{b}$ process.
The $gg\to tW\bar{b}$ process should pose no additional problems.
Therefore, this subtraction scheme paves the way to perform a NLO calculation of the $tWj$ production.

\acknowledgments

This work was supported in part by the National Science Foundation of China under grant No. 12275156, No. 12321005 and the Taishan Scholar Foundation of Shandong province (tsqn201909011).
The Feynman diagrams were drawn using {\tt TikZ-FeynHand}~\cite{Dohse:2018vqo}.
\appendix

\section{Momentum reshuffling}
\label{sec:MomResh}
In this appendix, we present the way of performing momentum reshuffling as in {\tt MC@NLO} and {\tt POWHEG-BOX} package. 
The motivation is to construct a phase space point for on-shell top quark production from the one for $tW\bar{b}$ production.
For clarity, the quantities with tilde denote the objects after momentum reshuffling.

The invariant mass of the $tW\bar{b}$ system is likely below the top quark pair threshold $2m_t$.
It is required that the partonic center-of-mass energy is reshuffled to be always larger than this threshold, i.e., 
\begin{align}
    \tilde{s}=s\left(\frac{2 m_t}{m_t+m_{W\bar{b}}}\right)^2,
\end{align}
where $m_{W\bar{b}}$ is the invariant mass of the $W\bar{b}$ system.
The reshuffled momenta of the initial-state partons are given by
\begin{align}
   \tilde{p}_1^{ \mu}=\sqrt{\tilde{s}}/2(1, \;0, \;0, \;1), \quad\quad \tilde{p}_2^{ \mu}=\sqrt{\tilde{s}}/2(1, \;0, \;0, \;-1).
    \label{eq: initalresh}
\end{align}
The rapidity $\eta$ of the initial state in the laboratory frame is not affected by the momentum reshuffling, 
and the momentum fractions of the initial-state partons read
\begin{align}
    \tilde{x}_1=\sqrt{ \tilde{\tau}}\;e^{\eta}, \quad\quad \tilde{x}_2=\sqrt{ \tilde{\tau}}\;e^{-\eta}
\end{align}
with \begin{align}
    \tilde{\tau}=\frac{\tilde{s}}{S},
\end{align}
where $\sqrt{S}$ denotes the collider energy. 

The momentum direction of the final-state top quark remains the same, with the polar and azimuthal angles given by
\begin{align}
    \theta_t=\arccos(p_t^3/|\vec{p}_t|), \quad\quad \phi_t=\arctan(p_t^2/p_t^1),
\end{align}
where $p_t^i$ is the $i$th component of $p_t$. 
The reshuffled momentum of the top quark can be written as
\begin{align}
    \tilde{p}_t^{ \mu}=\left(\sqrt{\tilde{s}}/2, \quad |\vec{\mathbf{p}}_t|\sin\theta_t\cos\phi_t, \quad |\vec{\mathbf{p}}_t|\sin\theta_t\sin\phi_t, \quad |\vec{\mathbf{p}}_t|\cos\theta_t\right),
    \label{eq: topresh}
\end{align}
with
\begin{align}
    |\vec{\mathbf{p}}_t|=\sqrt{\tilde{s}/4-m_t^2}.
\end{align}
Due to momentum conservation, the reshuffled momentum of the anti-top quark is 
\begin{align}
     \tilde{p}_{\bar t}^{ \mu}=\left(\sqrt{\tilde{s}}/2, \quad -|\vec{\mathbf{p}}_t|\sin\theta_t\cos\phi_t, \quad -|\vec{\mathbf{p}}_t|\sin\theta_t\sin\phi_t, \quad -|\vec{\mathbf{p}}_t|\cos\theta_t\right).
\end{align}

It is better to perform the momentum reshuffling of the $W$ boson and the anti-bottom quark  in their center-of-mass frame.
The momentum of the $W$ boson in this frame is obtained by
\begin{align}
    k_W^\mu=\Lambda^\mu_{\;\;\nu}(\vec{\beta}) \;p_W^\nu, \quad\quad
    \vec{\beta}=-\frac{\vec{p}_W+\vec{p}_{\bar b}}{p_W^0+p_{\bar b}^0},
\end{align}
where $\Lambda^\mu_{\;\;\nu}(\vec{\beta})$ is the Lorentz transformation matrix corresponding to the boost velocity $\vec{\beta}$.
The momentum direction is not changed, with the angles given by
\begin{align}
    \theta_W=\arccos(k_W^3/|\vec{k}_W|), \quad\quad \phi_W=\arctan(k_W^2/k_W^1).
\end{align}
The reshuffled momenta of the $W$ boson and the anti-bottom quark in their center-of-mass frame are represented by
\begin{align}
    \tilde{k}^{ \mu}_W&=( \tilde{k}^{0}_W,\quad |\vec{\mathbf{k}}_{\bar b}|\sin\theta_W\cos\phi_W,\quad 
    |\vec{\mathbf{k}}_{\bar b}|\sin\theta_W\cos\phi_W,\quad |\vec{\mathbf{k}}_{\bar b}|\cos\theta_W),\\
    \tilde{k}^{ \mu}_{\bar b}&=(\tilde{k}^{0}_{\bar b},\quad -|\vec{\mathbf{k}}_{\bar b}|\sin\theta_W\cos\phi_W,\quad -|\vec{\mathbf{k}}_{\bar b}|\sin\theta_W\cos\phi_W,\quad -|\vec{\mathbf{k}}_{\bar b}|\cos\theta_W)
\end{align}
with 
\begin{align}
    \tilde{k}^{0}_W=\frac{m_t^2+m_W^2}{2m_t}, \quad \tilde{k}^{0}_{\bar b}=|\vec{\mathbf{k}}_{\bar b}|=\frac{m_t^2-m_W^2}{2m_t}.
\end{align}
The momenta in the initial-state partonic center-of-mass frame can be obtained by performing a boost,
\begin{align}
    \tilde{p}_W^{\mu}=\Lambda^\mu_{\;\;\nu}(\vec{\beta}) \;\tilde{k}_W^{\nu}, \quad\quad 
    \tilde{p}_{\bar b}^{\mu}=\Lambda^\mu_{\;\;\nu}(\vec{\beta}) \; \tilde{k}_{\bar b}^{ \nu}
    \label{eq: Wbresh}
\end{align}
with $\vec{\beta}=\vec{\tilde{\mathbf{p}}}_{\bar t}/\tilde{p}^{ 0}_{\bar t}$.

Eqs. (\ref{eq: initalresh}), (\ref{eq: topresh}) and (\ref{eq: Wbresh}) collect all the reshuffled momenta we need to calculate the subtraction term.

\section{Correspondence between $\log \Delta$ terms and dimensional divergences }
\label{sec:OffvsOn}

\begin{figure}
    \centering
    \begin{minipage}{0.4\textwidth}
        \centering
        \begin{tikzpicture}
        \begin{feynhand}
        \vertex(dg)at(0.1,0);\vertex(dxg)at(0.1,-1.5);\vertex(tg)at(1.5,-1.5);\vertex(txg)at(1.5,0);\vertex(d)at(-1,0){$d$};\vertex(dx)at(-1,-1.5){$\bar{d}$};\vertex(bx)at(3,0){$\bar{b}$};\vertex(t)at(3,-1.5){$t$};\vertex(twb)at(2.2,0);\vertex(w)at(3,-0.8){$W$};\propag[fer](d)to(dg);\propag[fer](dg)to(dxg);\propag[fer](dxg)to(dx);\propag[fer](bx)to(twb);\propag[fer](d)to(dg);\propag[fer,blue](twb)to(txg);\propag[fer,blue](txg)to(tg);\propag[fer,blue](tg)to(t);\propag[glu](dg)to(txg);\propag[glu](dxg)to(tg);\propag[bos,red](twb)to(w);
        \end{feynhand}
    \end{tikzpicture}
    \caption*{(a)}
    \end{minipage}
    \begin{minipage}{0.4\textwidth}
        \centering
        \begin{tikzpicture}
        \begin{feynhand}
        \vertex(dg)at(0,0);\vertex(dxg)at(0,-1.5);\vertex(tg)at(1.4,-1.5);\vertex(txg)at(1.4,0);\vertex(d)at(-1.1,0){$d$};\vertex(dx)at(-1.1,-1.5){$\bar{d}$};\vertex(tx)at(2.5,0){$\bar{t}$};\vertex(t)at(2.5,-1.5){$t$};\propag[fer](d)to(dg);\propag[fer](dg)to(dxg);\propag[fer](dxg)to(dx);\propag[fer,blue](tx)to(txg);\propag[fer](d)to(dg);\propag[fer,blue](txg)to(tg);\propag[fer,blue](tg)to(t);\propag[glu](dg)to(txg);\propag[glu](dxg)to(tg);
        \end{feynhand}
    \end{tikzpicture}
    \caption*{(b)}
    \end{minipage}
    \caption{Sample Feynman diagrams appearing in the calculation of the $d\bar{d}\to\bar{b}Wt$ and $d\bar{d}\to\bar{t}t$ processes, respectively.}
    \label{fig:Box}
\end{figure}

As mentioned in the main text, the resonance subtraction term contains logarithms of $\Delta$ at the one-loop level.
Typical Feynman diagrams that contribute to such logarithms can be seen in figure \ref{fig:top_location}.
Part of these logarithms arise because $\Delta$ serves as a regulator of the infra-red divergence in the calculation of loop integrals.
Consequently, they are related to the divergences of the corresponding loop diagrams with an on-shell anti-top quark ($\Delta=0$) in dimensional regularization. 
For illustration, we show two interrelated Feynman diagrams in figure \ref{fig:Box}.
After expansion in $\Delta$, the result of the left diagram in figure \ref{fig:Box}, interfered with the LO amplitude, reads
\begin{align}
   \mathcal{M}_a  \mathcal{M}_{\rm LO}^* & = \frac{\alpha_s(2C_F-C_A)}{2\pi} \frac{B^{(2)}}{\Delta^2} \Bigg\{\frac{1}{2\epsilon^2}-\frac{1}{\epsilon}\bigg[\log\left(\frac{\mu^2}{-\Delta}\right)+\frac{1}{2}\log\left(\frac{\mu^2}{m_t^2}\right)\\
   &-2\log\left(\frac{\mu^2}{2\,p_{\bar d}\cdot p_t}\right)\bigg] 
    +\log\left(\frac{\mu^2}{m_t^2}\right)\log\left(\frac{\mu^2}{-\Delta}\right)-\log^2\left(\frac{\mu^2}{-\Delta}\right)\Bigg\}+\cdots,\nonumber
\end{align}
where we have written the divergences and $\log\Delta$ terms explicitly, and $p_{\bar d}$ and $p_t$ are the momenta of partons $\bar{d}$ and $t$, respectively.
In order to take the limit of $\Delta \to 0$,
we need to reproduce the full $\Delta$ dependence without expansion in $\epsilon$.
After observing the logarithmic structure in the above equation, we obtain
\begin{align}
   \mathcal{M}_a \mathcal{M}_{\rm LO}^* & =\frac{\alpha_s(2C_F-C_A)}{2\pi} \frac{B^{(2)}}{\Delta^2} \Bigg\{
   -\frac{1}{2\epsilon^2}\left(\frac{-\Delta}{\mu m_t}\right)^{-2\epsilon}
   \nonumber \\
   &+ \frac{1}{\epsilon^2}
   -\frac{1}{\epsilon}\bigg[\log\left(\frac{\mu^2}{m_t^2}\right) 
   -2\log\left(\frac{\mu^2}{2\,p_{\bar d} \cdot p_t}\right)\bigg] 
    \Bigg\}+\cdots.
\end{align}
Actually, we can consider the results in the first and second lines coming from the integrals in the soft and hard regions, respectively, as indicated by their typical scales.
The result in the first line is confirmed by direct computation of the soft loop integral in (\ref{eq:softintegral3}).
Now if we set $\Delta^{-2\epsilon}$ to 0 when taking the on-shell limit, we find full agreement with the divergence structure of the right diagram in figure \ref{fig:Box}, which is given by
\begin{align}
    \mathcal{M}_b^{\rm div} {\mathcal{M}^{d\bar{d}\to t\bar{t}}_{\rm LO}}^*
    &= |\mathcal{M}_{\rm LO}^{d\bar{d}\to t\bar{t}} |^2 \;\frac{\alpha_s(2C_F-C_A)}{2\pi}\left\{\frac{1}{\epsilon^2}-\frac{1}{\epsilon}\left[\log\left(\frac{\mu^2}{m_t^2}\right)-2\log\left(\frac{\mu^2}{2\,p_{\bar d} \cdot p_t}\right)\right]\right\}.
    \label{eq:ttBox}
\end{align}

\section{Soft loop integrals with an offshell propagator}
\label{sec:integrals}

\begin{figure}
    \centering
    \begin{minipage}{0.4\textwidth}
        \centering
        \begin{tikzpicture}
        \begin{feynhand}
        \vertex(dg)at(0.2,0.5);\vertex(dxg)at(0.2,-1.5);\vertex(tg)at(1.5,-1.5);\vertex(txg)at(1.5,0.5);\vertex(d)at(-1,0.5){$d(p_1)$};\vertex(dx)at(-1,-1.5){$\bar{d}(p_2)$};\vertex(bx)at(3,0.5){$\bar{b}(p_3)$};\vertex(t)at(3,-1.5){$t(p_5)$};\vertex(twb)at(1.5,-0.5);\vertex(w)at(3,-0.5){$W(p_4)$};\propag[fer](d)to(dg);\propag[fer](dg)to(dxg);\propag[fer](dxg)to(dx);\propag[fer](bx)to(txg);\propag[fer](txg)to(twb);\propag[fer,blue](twb)to(tg);\propag[fer,blue](tg)to(t);\propag[glu,red](txg)to[edge label=$l$](dg);\propag[glu](dxg)to(tg);\propag[bos,red](twb)to(w);
        \end{feynhand}
    \end{tikzpicture}
    \caption*{(a)}
    \end{minipage}
    \begin{minipage}{0.4\textwidth}
        \centering
        \begin{tikzpicture}
        \begin{feynhand}
        \vertex(ddg)at(0,0);\vertex(ttg1)at(1,0);\vertex(bbg)at(2,0);\vertex(b)at(3,0){$\bar{b}(p_3)$};\vertex(d1)at(-1,1){$d(p_1)$};\vertex(d2)at(-1,-1){$\bar{d}(p_2)$};\vertex(twb)at(1.5,1);\vertex(w)at(3,1){$W(p_4)$};\vertex(ttg2)at(1.5,-1);\vertex(t)at(3,-1){$t(p_5)$};\propag[fer](d1)to(ddg);\propag[fer](ddg)to(d2);\propag[fer](b)to(bbg);\propag[fer](bbg)to(twb);\propag[fer,blue](twb)to(ttg1);\propag[fer,blue](ttg1)to(ttg2);\propag[fer,blue](ttg2)to(t);\propag[glu](ddg)to(ttg1);\propag[glu,red](ttg2)to[edge label=$l$](bbg);\propag[bos,red](twb)to(w);
        \end{feynhand}
    \end{tikzpicture}
    \caption*{(b)}
    \end{minipage}
    \begin{minipage}{0.4\textwidth}
        \centering
        \begin{tikzpicture}
        \begin{feynhand}
        \vertex(dg)at(0.2,0.5);\vertex(dxg)at(0.2,-1.5);\vertex(tg)at(1.5,-1.5);\vertex(txg)at(1.5,0.5);\vertex(d)at(-1,0.5){$d(p_1)$};\vertex(dx)at(-1,-1.5){$\bar{d}(p_2)$};\vertex(bx)at(3,0.5){$\bar{b}(p_3)$};\vertex(t)at(3,-1.5){$t(p_5)$};\vertex(twb)at(2,0.5);\vertex(w)at(3,-0.5){$W(p_4)$};\propag[fer](d)to(dg);\propag[fer](dg)to(dxg);\propag[fer](dxg)to(dx);\propag[fer](bx)to(twb);\propag[fer](d)to(dg);\propag[fer,blue](twb)to(txg);\propag[fer,blue](txg)to(tg);\propag[fer,blue](tg)to(t);\propag[glu,red](txg)to[edge label=$l$](dg);\propag[glu](dxg)to(tg);\propag[bos,red](twb)to(w);
        \end{feynhand}
    \end{tikzpicture}
    \caption*{(c)}
    \end{minipage}
    \begin{minipage}{0.4\textwidth}
        \centering
        \begin{tikzpicture}
        \begin{feynhand}
        \vertex(ddg)at(0,0);\vertex(ttg1)at(0.7,0);\vertex(twb)at(1.9,1);\vertex(w)at(3,0){$W(p_4)$};\vertex(d1)at(-1,1){$d(p_1)$};\vertex(d2)at(-1,-1){$\bar{d}(p_2)$};\vertex(ttg3)at(1.5,1);\vertex(b)at(3,1){$\bar{b}(p_3)$};\vertex(ttg2)at(1.5,-1);\vertex(t)at(3,-1){$t(p_5)$};\propag[fer](d1)to(ddg);\propag[fer](ddg)to(d2);\propag[fer](b)to(twb);\propag[fer,blue](twb)to(ttg3);\propag[fer,blue](ttg3)to(ttg1);\propag[fer,blue](ttg1)to(ttg2);\propag[fer,blue](ttg2)to(t);\propag[glu](ddg)to(ttg1);\propag[glu,red](ttg2)to[edge label=$l$](ttg3);\propag[bos,red](twb)to(w);
        \end{feynhand}
    \end{tikzpicture}
    \caption*{(d)}
    \end{minipage}
    \caption{The Feynman diagrams containing soft loop integrals. The momentum of the soft gluon is denoted by $l$.}
    \label{fig:resdiagram}
\end{figure}

In our case, there are four types of soft loop integrals contributing to the double resonant part in the one-loop amplitude.
The Feynman diagrams containing such integrals are shown in figure \ref{fig:resdiagram}.
The analytical results of these soft loop integrals are given by
\begin{align}
  I_s^{(a)} =& \frac{\mu^{4-D}}{i\pi^{D/2}r_\Gamma}\int d^Dl \frac{1}{(l^2+i0)\,(-2l \cdot p_3+i0)\,[-2l \cdot p_{34}+\Delta+i0]\,(-2l \cdot p_1+i0)} \notag\\
    =&-\frac{1}{2(p_1 \cdot p_3)\Delta}\left(\frac{-\Delta-i0}{\mu\,m_t}\right)^{-2\epsilon}\Bigg[\frac{1}{\epsilon^2}-\frac{1}{\epsilon}\log\left(\frac{m_t^2(p_1 \cdot p_3)}{2\,(p_1 \cdot p_{34})(p_3 \cdot p_{34})}\right) \notag\\
    &+\frac{1}{2}\log^2\left(\frac{m_t^2(p_1 \cdot p_3)}{2\,(p_1 \cdot p_{34})(p_3 \cdot p_{34})}\right)+\text{Li}_2\left(1-\frac{m_t^2(p_1 \cdot p_3)}{2\,(p_1 \cdot p_{34})(p_3 \cdot p_{34})}\right)+\frac{\pi^2}{2}\Bigg]
\end{align}
in the case of $p_1^2=p_3^3=0$,
and
\begin{align}
     I_s^{(b)} = &\frac{\mu^{4-D}}{i\pi^{D/2}r_\Gamma}\int d^Dl\frac{1}{(l^2+i0)\,(-2l \cdot p_3+i0)\,[-2l\cdot p_{34}+\Delta+i0]\,(2l \cdot p_5+i0)} \notag\\
    =&\frac{1}{2(p_3 \cdot p_5)\Delta}\left(\frac{-\Delta-i0}{\mu\,m_t}\right)^{-2\epsilon}\Bigg\{\frac{1}{2\epsilon^2}+\frac{1}{\epsilon}\left[\log\left(\frac{p_3 \cdot p_{34}}{p_3 \cdot p_5}\right)+i\pi\right] \notag\\
    &-\log^2\left(\frac{1-\beta_t}{1+\beta_t}\right)-\text{Li}_2\left(1-\frac{1-\beta_t}{1+\beta_t}\frac{p_3 \cdot p_{34}}{p_3 \cdot p_5}\right)-\text{Li}_2\left(1-\frac{1+\beta_t}{1-\beta_t}\frac{p_3 \cdot p_{34}}{p_3 \cdot p_5}\right) \notag\\
    &+\frac{5\pi^2}{4}+2\pi i\log\left(\frac{1+\beta_t}{1-\beta_t}-\frac{p_3 \cdot p_{34}}{p_3 \cdot p_5}\right)\Bigg\}
\end{align}
in the case of $p_3^2=0, p_5^2=m_t^2$.
The other two soft loop integrals are given by
\begin{align}
     I_s^{(c)} = &\frac{\mu^{4-D}}{i\pi^{D/2}r_\Gamma}\int d^Dl\frac{1}{(l^2+i0)\,[-2l \cdot p_{34}+\Delta+i0]\,(-2l \cdot p_1+i0)} \notag\\
    =&\frac{1}{2(p_1 \cdot p_{34})}
    \left(\frac{-\Delta-i0}{\mu\,m_t}\right)^{-2\epsilon}
    \left(\frac{1}{2\epsilon^2}+\frac{\pi^2}{4}\right)
    \label{eq:softintegral3}
\end{align}
and
\begin{align}
     I_s^{(d)} = &\frac{\mu^{4-D}}{i\pi^{D/2}r_\Gamma}\int d^Dl\frac{1}{(l^2+i0)\,[-2l \cdot p_{34}+\Delta+i0]\,(2l \cdot p_5+i0)} \notag\\
    =&\frac{1-\beta_t^2}{\beta_t} \frac{1}{4\,m_t^2}
    \left(\frac{-\Delta-i0}{\mu\,m_t}\right)^{-2\epsilon}
    \Bigg\{
    -\frac{1}{\epsilon} \left[\log \left(\frac{1-\beta_t}{1+\beta_t}\right) +i \pi \right]
    +\log^2 \left(\frac{1-\beta_t}{1+\beta_t}\right)  
    \nonumber \\  & 
     +\mathrm{Li}_2\left(\frac{4\beta_t}{(1+\beta_t)^2} \right) - \pi^2 - 2  \pi i \log\left(\frac{4\beta_t}{1-\beta_t^2}\right)   
    \Bigg\}.
\end{align}
The normalization factor in the above equations is defined by
\begin{align}
    r_\Gamma & =\frac{\Gamma^2(1-\epsilon)\Gamma(1+\epsilon)}{\Gamma(1-2\epsilon)}.
\end{align}

\bibliography{biblio}
\bibliographystyle{JHEP}

\end{document}